\RequirePackage{ifpdf}
\pdfoutput=1
\documentclass{JINST}
\usepackage{booktabs,array,multirow}
\newcommand{\1}[1]{\, \mathrm{#1}}
\newcommand{\n}[1]{\mathrm{#1}}    
\newcommand{\Th}{$^{228}$Th}
\newcommand{\Ra}{$^{224}$Ra}
\newcommand{\Rn}{$^{220}$Rn}
\newcommand{\Pb}{$^{212}$Pb}

\title{A \boldmath $^{220}$Rn source for the calibration of low-background experiments}

\author{R.~F.~Lang$^{a}$, A.~Brown$^{a,b}$, E.~Brown$^{c,d}$, M.~Cervantes$^{a}$, S.~Macmullin$^{a}$, D.~Masson$^{a,}$\thanks{Corresponding author.}, J.~Schreiner$^{e}$, H.~Simgen$^{e}$ \\
\llap{$^a$}Department of Physics and Astronomy, Purdue University,\\ West Lafayette, IN, USA \\
\llap{$^b$}Nikhef and the University of Amsterdam,\\ Science Park, Amsterdam, Netherlands \\
\llap{$^c$}Institut f\"ur Kernphysik,\\ Wilhelms-Universit\"at M\"unster, M\"unster, Germany \\
\llap{$^d$}Department of Physics, Applied Physics and Astronomy,\\ Rensselaer Polytechnic Institute, Troy, NY, USA \\
\llap{$^e$}Max-Planck-Institut f\"ur Kernphysik,\\ Heidelberg, Germany \\

E-mail: \email{dmasson@purdue.edu}}

\abstract{We characterize two $40\1{kBq}$ sources of electrodeposited \Th~for use in low-background experiments. The sources efficiently emanate \Rn, a noble gas that can diffuse in a detector volume. \Rn~and its daughter isotopes produce $\alpha$-, $\beta$-, and $\gamma$-radiation, which may used to calibrate a variety of detector responses and features, before decaying completely in only a few days. We perform various tests to place limits on the release of other long-lived isotopes. In particular, we find an emanation of $<0.008\1{atoms/min/kBq}$ (90\% CL) for~\Th~and $(1.53\pm0.04)\1{atoms/min/kBq}$ for~\Ra. The sources lend themselves in particular to the calibration of detectors employing liquid noble elements such as argon and xenon. With the source mounted in a noble gas system, we demonstrate that filters are highly efficient in reducing the activity of these longer-lived isotopes further. We thus confirm the suitability of these sources even for use in next-generation experiments, such as XENON1T/XENONnT, LZ, and nEXO.}


\begin{document}
\maketitle
\flushbottom

\section{Introduction}\label{sec:introduction}

Various low-background experiments are searching for rare events such as neutrinoless double-beta decay~\cite{Pandola:2014naa} or dark matter scatters~\cite{Undagoitia:2015gya}. Time projection chambers (TPCs) using liquid noble elements such as argon and xenon are at the forefront of these investigations~\cite{Albert:2015ekt,Aprile:2013doa,Akerib:2015rjg,Amaudruz:2014nsa,Calvo:2015uln,Agnes:2015ftt}. As these detectors become large, new challenges arise in accurately calibrating their response throughout the detection volume. For example, the XENON1T detector~\cite{Aprile:2015uzo} is a TPC $1\1{m}$ in diameter and in height, much larger than e.g. the $70\1{mm}$ mean free path of the $1.3\1{MeV}$ $\gamma$ from $^{60}$Co. Thus, calibration of the entire fiducial volume with external sources is no longer feasible.

One solution is to directly mix a radioactive isotope into the liquid noble element. In liquid xenon, $^{83m}$Kr has been proposed~\cite{Hannen:2011,Manalaysay:2009yq,Kastens:2010} and used~\cite{Akerib:2013tjd} to achieve a low-energy $\gamma$-line calibration throughout the detector. Tritiated methane has been used as a low-energy beta source~\cite{Akerib:2013tjd}. While this source has the advantage that all decays are in the low-energy range interesting to a dark matter search, it suffers from the disadvantage that the activity does not decay by itself but must be extracted from the liquid target by a hot zirconium getter in a xenon recirculation loop~\cite{Akerib:2015wdi}. This introduces an additional time constant that will get longer as the size of these detectors increases.

\begin{figure}[htbp]
\centering
\includegraphics[trim = 5 0 50 15, clip = true,width = 0.8\columnwidth]{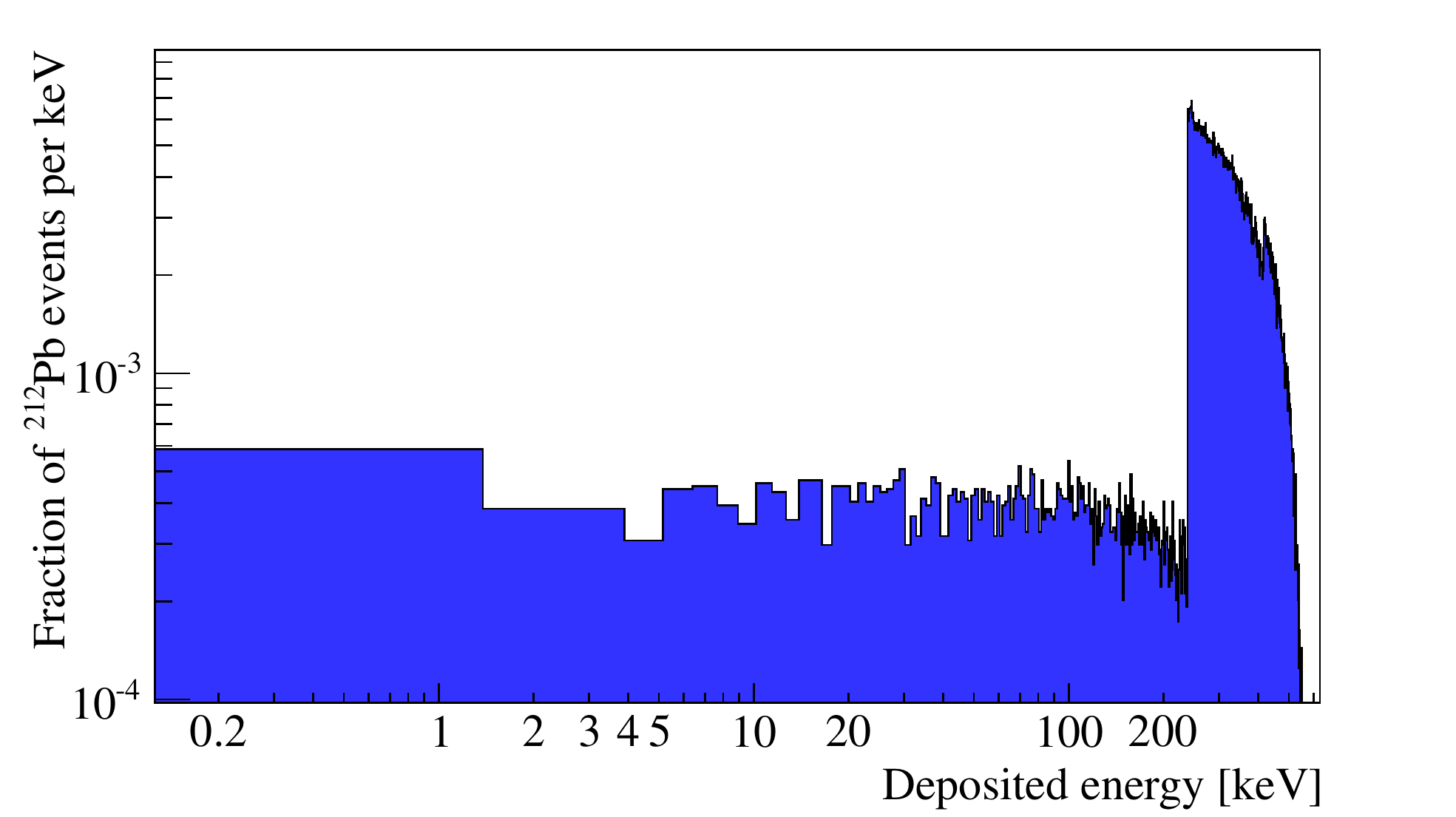}
\caption{A simulation of the $\beta$-spectrum of \Pb. The feature at 240 keV is due to decay modes with an associated $\gamma$ that adds to the energy observed in the decay.}
\label{fig:pb212spectrum}
\end{figure}

Here, we study a source containing \Th~that emanates \Rn~($T_{1/2}=56\1{seconds}$), which in turn can be mixed into the liquid target. The \Rn~decay chain is versatile, producing $\alpha$-, $\beta$-, and $\gamma$-radiation that makes this source interesting for a wide range of applications: The source is the ideal calibration source for intrinsic $^{222}$Rn backgrounds that are notoriously difficult to control in low-background experiments. (2) High-energy alphas~\cite{WeberM:2013,Albert:2015vma} and $^{212}$Bi/$^{212}$Po decays~\cite{Bellini:2012qg} can be used to accurately understand intrinsic backgrounds. The relatively short half-life of $^{216}$Po (145~ms) allows for the measurement of currents or liquid flows within the bulk of the detector volume. Using position reconstruction algorithms that yield the positions of the decays of \Rn~and $^{216}$Po together with the time difference between these two decays, allows for a measurement of the drift velocity of polonium ions~\cite{Albert:2015vma}. Thus, currents within the bulk of the detector can be mapped to identify possible dead regions that do not participate in the recirculation of the active volume through purification systems. Additionally, the high-energy $\alpha$ lines can be used to probe regions in a detector with poor light or charge collection efficiency. Further, the $^{208}$Tl decay produces a $2.6\1{MeV}$ $\gamma$-line, very close to the Q-value of $^{136}$Xe double-beta decay. This line will be accompanied by the $\beta$ energy and usually by simultaneous $\gamma$-lines (mostly $580\1{keV}$), which results in steps in the calibration spectrum (e.g. at $\approx 3.2\1{MeV}$) above the $^{136}$Xe Q-value. Finally, $\beta$-decays of the chain that go to the ground state can be used to calibrate the low-energy response of dark matter detectors. While the high Q-value of the $^{212}$Bi $\beta$-decay (2.2~MeV) and short life-time of the daughter $^{212}$Po (300~ns) render that decay unsuitable for this purpose, the $\beta$-decay of \Pb~(12.3\% branching ratio to ground state, Q-value 560~keV) is very suitable for this task, as shown in Figure~\ref{fig:pb212spectrum}.

A major advantage of our source is that the time scale of the \Rn~decay chain is dominated by the relatively short half-life of \Pb~($T_{1/2}=10.66\1{hours}$). Thus, the introduced activity can completely decay away within a few days, making this source useful even for the largest anticipated liquid detectors~\cite{Baudis:2012,Akerib:2015cja,Franco:2015pha}. As both \Th~($T_{1/2}=1.9\1{years}$) and \Ra~($T_{1/2}=3.6\1{days}$) have much longer half-lives, emanation of these isotopes from the source must be limited. Also, isotopic contaminations with $^{230}$Th in the $^{228}$Th source itself can lead to the emanation of $^{222}$Rn ($T_{1/2}=3.8\1{days}$) which has to be avoided.

\begin{figure}[htbp]
\centering
\includegraphics[trim = 5 10 50 0, clip = true,width = 0.8\columnwidth]{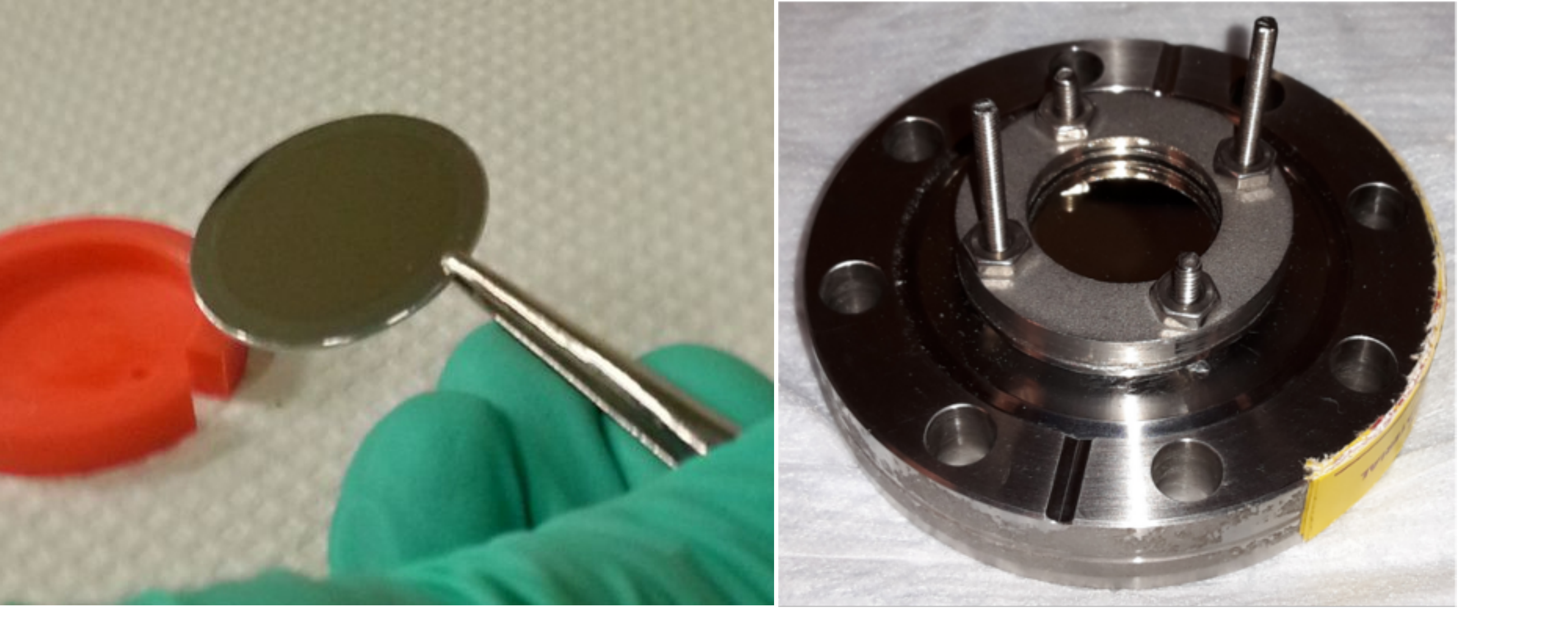}
\caption{\Th~is deposited onto a $30\1{mm}$ stainless steel disc (left). It is held in a simple emanation vessel (right) for mounting on a noble gas system.}
\label{fig:th228source}
\end{figure}

Here, we use a variety of methods to derive limits on the release of long-lived isotopes, and demonstrate that these sources are suitable for the calibration of even next-generation low-background experiments. Open \Rn~sources were produced by electroplating thorium nitrate, $\n{Th}(\n{NO}_3)_4$, onto the center of a $30\1{mm}$ diameter stainless steel disk in a bath of 1M nitric acid ($\n{HNO}_3$). A ring of width 2.5~mm around the edge of the disk was left for mounting purposes. The activity was $40\1{kBq}$ as of March, 2015. Each source is held in a small stainless steel vessel to attach it to a noble gas recirculation system using 1/2"~VCR piping, see figure~\ref{fig:th228source}.

\section{$\gamma$-Measurements of Filter Deposition}
\label{sec:tuv}

The source was tested
for the release of \Ra~and \Th~using a standard procedure. Nitrogen was flushed through the source vessel for 96 hours. A filter of type ML050/0 was mounted inline $18\1{cm}$ after the source, containing a filter paper on which any released radionuclides could be deposited. This filter paper was then tested for $\gamma$-activity with a high-purity germanium detector. A first measurement was made immediately after exposure, and a second measurement a week later, see Table~\ref{tab:tuv}.

\begin{table}[htb]
\centering
\caption{Measurements of radionuclide release from the \Rn~source collected in filter paper.}
\label{tab:tuv}
\renewcommand{\arraystretch}{1.2}
\begin{tabular}[c]{|llcc|}
\hline\hline
\multicolumn{2}{|l}{Measurement} & 1 & 2 \\
\multicolumn{2}{|l}{Time after exposure} & Immediate & 1 week \\
\multicolumn{2}{|l}{Livetime} & $375\1{s}$ & $11500\1{s}$ \\ \hline
\multirow{5}{*}{Activity/Bq}
& \Th & $<35$ & $<1.97$ \\
& \Ra & $<6$ & $<0.61$ \\
& \Pb & $87\pm11$ & $<0.07$ \\
& $^{212}$Bi & $84\pm34$ & $<0.68$ \\
& $^{208}$Tl & $28\pm5$ & $<0.07$ \\
\hline\hline
\end{tabular}
\end{table}

Both the exposure time and the time between measurements are significant compared to the half-life of \Ra. We account for both the decay during these intervals as well as the production of \Ra~from the decay of \Th. The week between the two measurements is more than 15 half-lives of \Pb, hence any recorded activity of its daughters $^{212}$Bi or $^{208}$Tl in the second measurement would have been from the decay of \Ra, not any initial population of \Pb. We use the lowest measurement (here, \Pb) to constrain the release of \Ra~from the source to $<0.43\1{atoms/s}$ and that of \Th~to or $<22\1{atoms/s}$. Scaling these values for the activity of the source yields a stray emanation of $<0.66\1{atoms/min/kBq}$ \Ra~and $<34\1{atoms/min/kBq}$ \Th. We choose these units (atoms/min/kBq) to account for the decay of the sources over the time the various measurements were made, and to allow comparisons between the sources.

The obvious limitation of this measurement is that there may have been radium or thorium released by the source but not caught by the filter, in which case the given limits must be scaled by the efficiency of the filter. Additionally, any thorium or radium plated out on the pipes connecting the source vessel and the filter would not show up in this measurement.

\begin{figure}[htb]
\centering
\includegraphics[trim = 15 0 50 20, clip = true,width = 0.8\columnwidth]{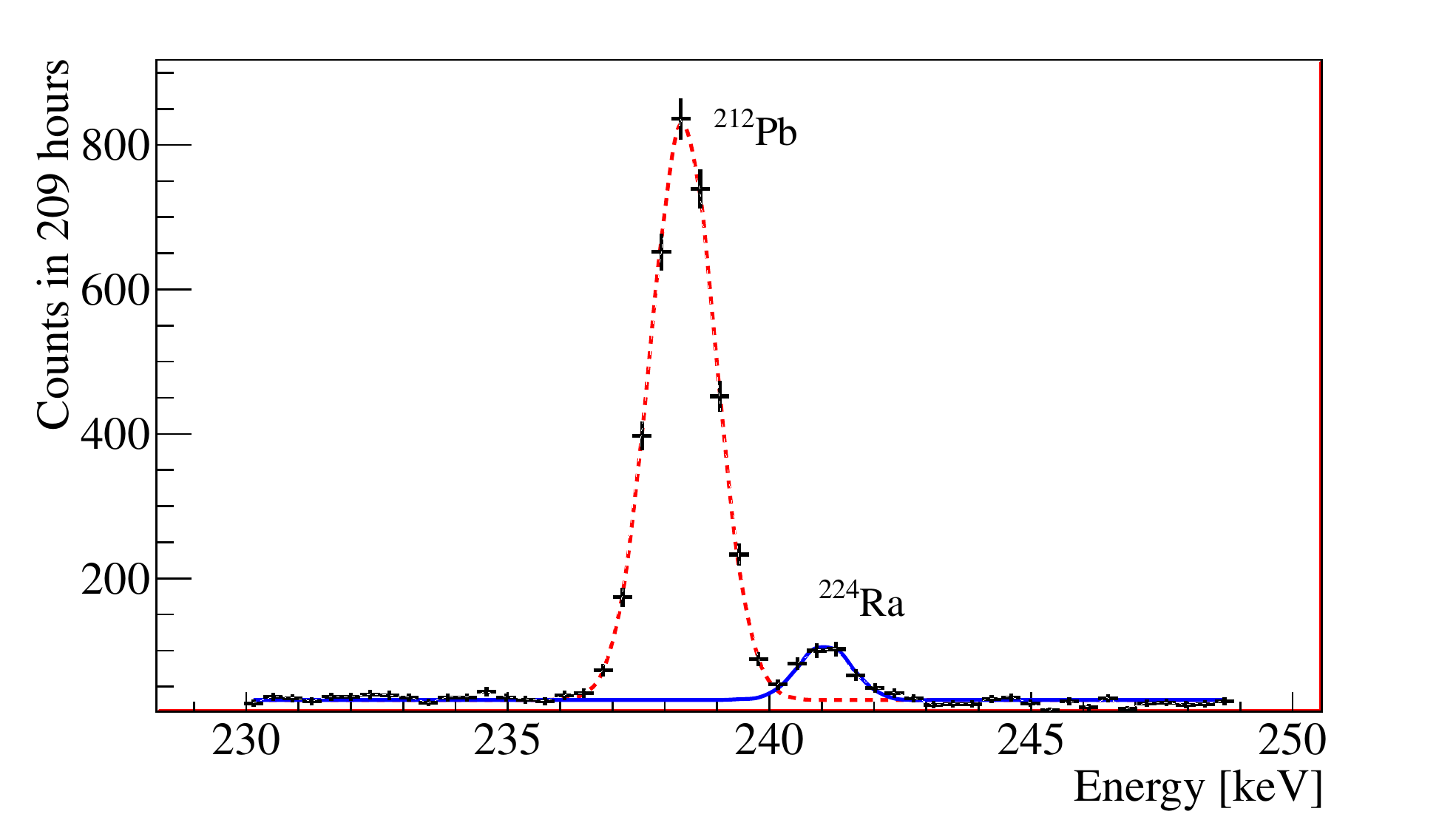}
\caption{Gamma spectrum taken of the filter 7.2 days after exposure showing the \Pb~line at 238.6 keV and the \Ra~line at 241.0 keV. The dashed and solid lines are from a fit to the data.}
\label{fig:gammaspectrum}
\end{figure}

A similar but more sensitive experiment was performed by pumping nitrogen at 1 standard liter per minute (slpm) for 9 days in a closed loop through the \Rn~source vessel. The source was followed by a MILLEX-FG 50 filter at a distance of 8~cm, containing a $0.2\1{\mu m}$ PTFE filter membrane. The exposure time brought any released \Ra~nearly into equilibrium on the filter and gave potential \Th~more time to deposit itself. After exposure, the filter membrane was tested for $\gamma$-activity with high-purity germanium detectors~\cite{Budjas:1,Budjas:2}. A spectrum of these measurements is shown in figure~\ref{fig:gammaspectrum}. A simulation of the germanium crystal detector geometry performed in GEANT4~\cite{Geant4} indicated an efficiency of 6.5\% at 240 keV.

A week after exposure most of the \Pb~has decayed and measurements become sensitive to \Ra. A measurement with a livetime of 208 hours yielded a \Ra~activity of $(1.0\pm0.3)\1{Bq}$ on the filter at the end of exposure. A second measurement was done forty-two days after exposure, at which point the \Ra~deposited on the filter should have decayed to $3\times10^{-4}$ of the initial population. Thus, any measured \Ra~activity could be attributed to residual \Th. We calculate upper limits (90\% CL) of 2.5 mBq for \Th~and 2.4 mBq for \Ra~on the filter. These activities convert to emanation rates of $(1.9\pm0.6)\1{atoms/min/kBq}$ \Ra~and $<0.4\1{atoms/min/kBq}$ \Th.

\section{$^{222}$Rn Emanation Measurement}

Potential traces of $^{226}$Ra or $^{230}$Th in the \Rn~source would lead to a non-negligible emanation of the long-lived radon isotope $^{222}$Rn, which must be avoided. We applied ultra-low background proportional counters as described in~\cite{zuzel_llrmt} to measure directly the $^{222}$Rn emanation rate of the source. For this purpose the source was connected to a 1 liter stainless steel buffer volume, separated by a valve. The setup was evacuated and the valve was opened such that \Rn~and $^{222}$Rn from the source could emanate into the buffer volume. After some days the valve was closed and the buffer volume was separated from the source. When all \Rn~had decayed, the $^{222}$Rn was extracted from the buffer volume and filled to a proportional counter in which the alpha decays of $^{222}$Rn and its daughters were counted.

We repeated the measurement three times with a small modification in the the third measurement: Instead of emanating into vacuum, we filled 1.5 bar of helium in the buffer volume to check whether there is a difference between radon emanation into gas and into vacuum. It turned out that this is not the case as all three measurements are in good agreement and compatible with zero. The combined result is a $^{222}$Rn emanation rate from the \Rn~source of $<55~\mu$Bq.

\section{Si PIN Diode Measurements}
\label{sec:diode}
A direct measurement of the \Rn~source was performed with a custom-developed radon monitor. It consists of a 3-liter vacuum-tight stainless steel vessel containing a 2 cm square windowless Si PIN diode from Hamamatsu. A high voltage of 1.5 kV collects the charged ions resulting from the decays of \Rn~onto the surface of the diode, where the $^{216}$Po decay can be detected with an efficiency of about 35\%. The radon detector was calibrated using a $^{226}$Ra solution with a known activity of $(25\pm1)\1{Bq}$ by bubbling nitrogen through the solution, thus obtaining a known activity of $^{222}$Rn.

The \Rn~source was placed directly inside the radon monitor, which was filled with air, and \Rn~was allowed to reach equilibrium. Assuming the same collection efficiency of $^{222}$Rn and \Rn, an emanation of $(1750\pm50)\1{Bq}$ \Rn~was determined. After 10~days, the \Rn~source was removed from the radon monitor and the vessel evacuated, and all collected ions on the surface of the Si PIN diode were left to decay. Six days after the removal of the source, all \Rn~activity would be due to emanated \Ra~collected on the surface of the diode. The emanated \Ra~activity was calculated to be $(2.1\pm0.7)\1{Bq}$ and the \Th~activity as $<50\,\mu\n{Bq}$ at the point when the source was removed, which corresponds to a \Ra~emanation rate of $(3.9\pm1.3)\1{atoms/min/kBq}$ and a \Th~rate of $<0.008\1{atoms/min/kBq}$.

\begin{figure}[htb]
\centering
\includegraphics[trim = 5 5 55 20, clip = true,width = 0.8\columnwidth]{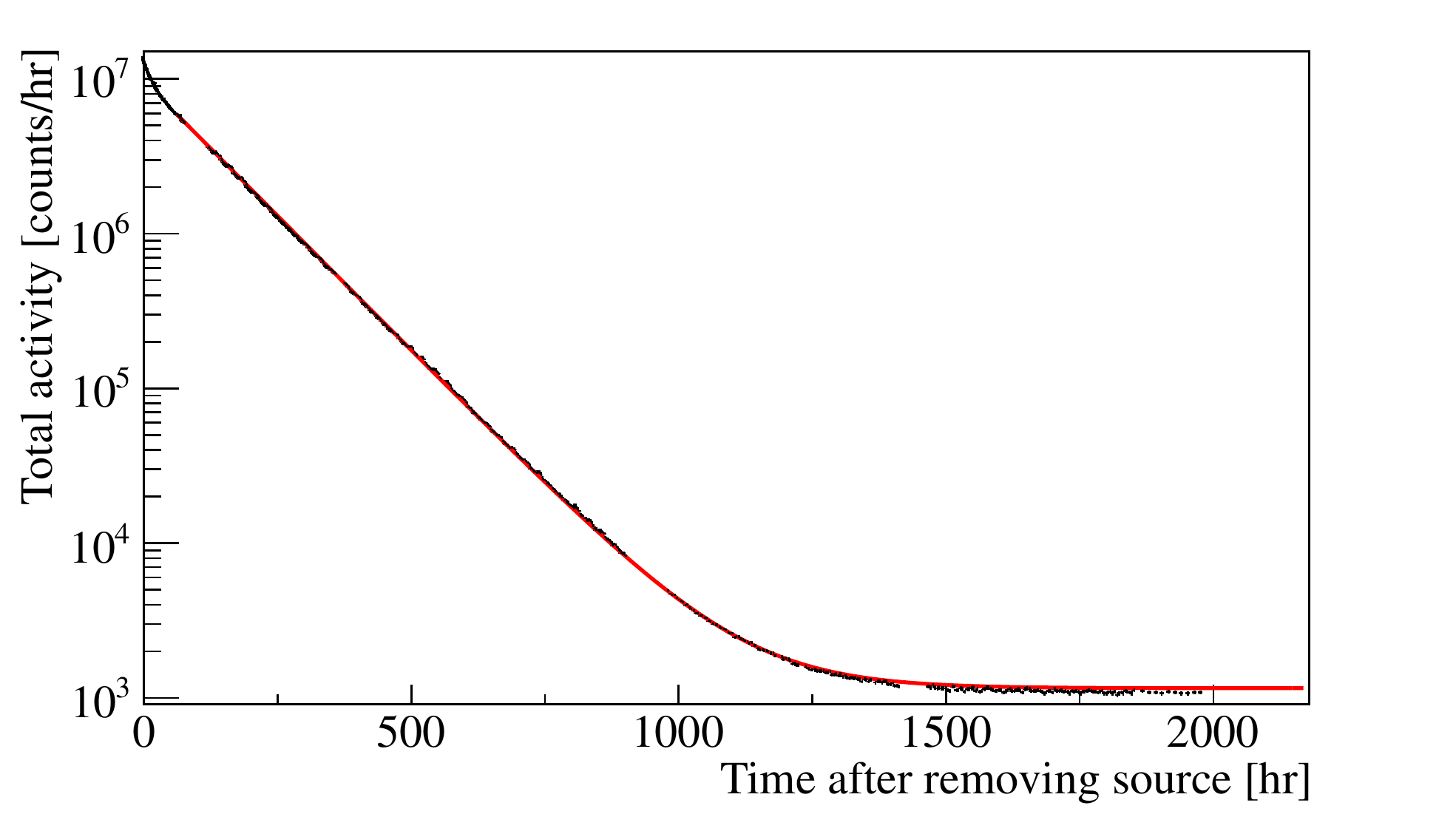}
\caption{Total activity in the Si PIN diode above 1 MeV. The data points include statistical error bars, barely visible on this scale. The line is the fit of an exponential plus constant baseline.}
\label{fig:pindiodespectrum}
\end{figure}

To further determine levels of emanation, the source was placed directly facing a Hamamatsu windowless Si PIN diode that acts as an alpha spectrometer~\cite{Bray}. The source and diode were placed in a small vessel separated by only $8\1{mm}$. The vessel was then evacuated, and the source was left for five days to deposit material onto the surface of the diode. The source was then removed and the atoms deposited onto the diode were left to decay under vacuum. Figure~\ref{fig:pindiodespectrum} shows the total activity in the diode above $1\1{MeV}$ decaying away after the removal of the source.

An exponential decay plus a constant baseline is fit to this data, yielding a decay constant in agreement with the accepted half-life of \Ra, and a residual background of $(1128\pm3)\1{hr^{-1}}$. Extrapolating the curve backwards and scaling for the fraction of total activity that is \Ra, we can calculate the emanation rate of \Ra~onto the surface of the diode to be $(924.0\pm0.3)\1{s^{-1}}$. Two months after removing the source, the initial population of \Ra~will have decayed away, so the background value inferred from the fit is due to a combination of intrinsic backgrounds (here, measured to be negligible) and released \Th. By measuring the activity in the relevant portion of the spectrum, we find a \Th~activity of $(0.097\pm0.003)\1{Bq}$, or $(8.4\pm0.3)\times10^6\1{atoms}$. Given the exposure time, this corresponds to a \Th~emanation rate of $(18.3\pm0.6)\1{s^{-1}}$ onto the surface of the diode. Simulation with GEANT4 yields a geometric efficiency for deposition of 0.13, which allows us to scale the emanation rates of the source to $(4.5\pm0.2)\1{atoms/s/kBq}$ \Th~and $(282.1\pm0.1)\1{atoms/s/kBq}$ \Ra.

\section{$\gamma$-Spectroscopy of Pipe Contamination}
\label{sec:flush}

A test of radium plate-out was performed by flushing argon from a high-pressure tank through a pressure regulator, the source vessel, and then a copper pipe of $6\1{mm}$ diameter and $50\1{cm}$ length. The argon flow averaged $6\1{slpm}$ though with sizeable fluctuations. After 41~hours of flushing, the copper pipe was cold-welded shut at both ends to seal in any materials deposited on the inner surface, and swiftly transported for measurement using our low-background germanium counters. Several measurements were done over the course of about two months to determine the $\gamma$-activity of the copper pipe. The results of the measurements are given in Table~\ref{tab:flush_meas}.
\begin{table}[htb]
\centering
\caption{Results of the measured $\gamma$-activity in the cold-welded copper pipe after flushing argon through source and pipe.}
\label{tab:flush_meas}
\renewcommand{\arraystretch}{1.2}
\begin{tabular}{|llccc|}
\hline\hline
\multicolumn{2}{|l}{Measurement} & 1 & 2 & 3 \\
\multicolumn{2}{|l}{Time after exposure} & 10 hours & 7 days & 68 days \\
\multicolumn{2}{|l}{Livetime} & 76733 s & 897820 s & 357802 s \\ \hline
\multirow{2}{*}{Activity/Bq}
& \Ra & N/A & $0.25\pm0.03$ & $<0.088$ \\
& \Pb & $13.8\pm0.1$ & $0.215\pm0.005$ & $<0.088$ \\
\hline\hline
\end{tabular}
\end{table}

The interpretation of these data is done in a very similar way to that presented in the previous section. The first measurement started approximately 1 half-life of \Pb~after exposure, so any amount deposited on the pipe would still be observable. By the time the second measurement was started, a week had passed ($>$15 half-lives), so any initial \Pb~would have decayed to a negligible amount, making the measurement sensitive to potential \Pb~from the decay of \Ra. The interval between exposure and the third measurement is many half-lives of \Ra, making the measurement sensitive to potential contamination from the parent \Th~in the copper pipe.

\begin{table}[htb]
\centering
\caption{Calculated values and limits on \Ra~and \Th~release from deposition in the copper pipe after argon flushing.}
\label{tab:flush_limits}
\renewcommand{\arraystretch}{1.2}
\begin{tabular}{|lcc|}
\hline\hline
Isotope & Activity after exposure & Emanation rate \\ \hline
\Pb & $(26.0\pm0.2)\1{Bq}$ & N/A \\
\Ra & $(0.95\pm0.11)\1{Bq}$ & $(1.53\pm0.04)\1{atoms/min/kBq}$ \\
\Th & $<0.093\1{Bq}$ & $<47\1{atoms/min/kBq}$ \\
\hline\hline
\end{tabular}
\end{table}

Indeed trace amounts of \Ra~appear to have come off the source in this experiment. While the overall activity is very small, it motivates the use of an additional filter just after the source vessel for the calibration of low background detectors.

\section{Measurements of Filter Efficacy}
\label{sec:filter}

In order to assess the performance of various filters in limiting the release of \Ra~from the \Rn~source, the source vessel was connected to a xenon gas system. Xenon gas was recirculated through various configurations involving the \Rn~source vessel, filters, and a Si PIN diode as an $\alpha$-spectrometer~\cite{Bray}. The source was exposed to the xenon gas stream for a few days, then bypassed, and the decaying activity monitored to measure any released radium. Two filter types were tested for their ability to remove radium from the gas stream. The first filter was a Swagelok F-series 0.5-micron sintered filter, the second was a Swagelok SCF-series ceramic filter.

\begin{figure}[htb]
\centering
\includegraphics[trim = 5 5 40 15, clip = true,width = 0.8\columnwidth]{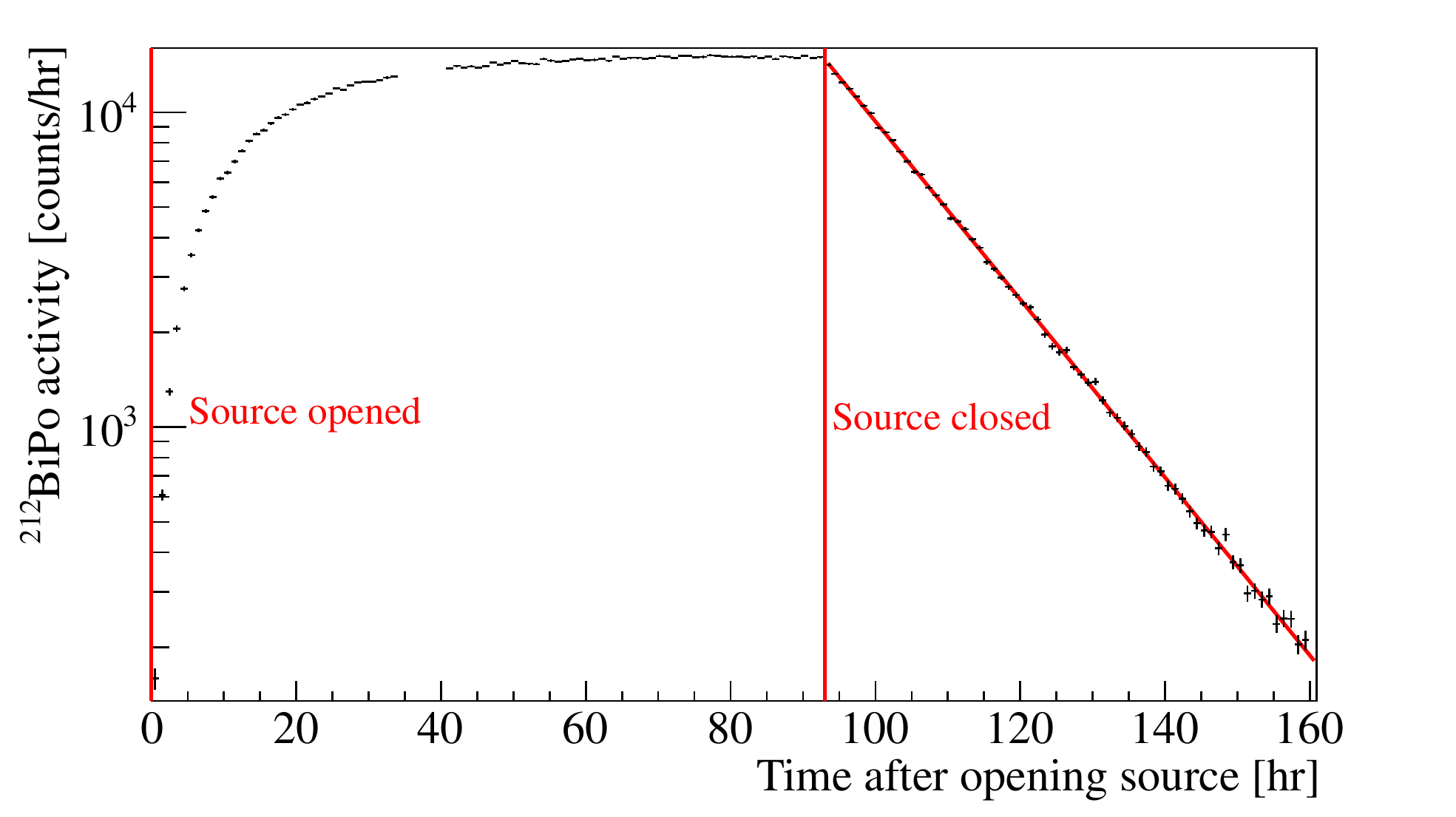}
\caption{Activity of $^{212}$BiPo in the Si PIN diode during and after flushing xenon through the ceramic filter. An exponential was fitted to the decaying part, yielding a half-life of $(10.64\pm0.05)\1{h}$ in excellent agreement with the \Pb~half-life of $(10.64\pm0.01)\1{h}$~\cite{Firestone}.}
\label{fig:bipo}
\end{figure}

The source vessel was connected directly to the sintered filter and then to the Si PIN diode with about $1\1{m}$ of 1/4" stainless steel pipe. For the ceramic filter, the extra piping was reduced to 8 cm. Xenon gas at $1\1{barg}$ was recirculated through the gas system at $5\1{slpm}$ for the sintered filter and $10\1{slpm}$ for the ceramic filter. After exposure with the sintered filter in line, the source and filter were bypassed and recirculation continued through the detector vessel for several weeks. After exposure with the ceramic filter, recirculation was stopped and the activity deposited in the detector vessel left to decay. Figure~\ref{fig:bipo} shows the activity of coincident $^{212}$Bi-$^{212}$Po (BiPo) activity in the Si PIN diode for the measurement of the ceramic filter from when the source was opened to three days after the source was closed, when the detector vessel was evacuated.

\begin{figure}[htb]
\centering
\includegraphics[trim = 10 5 40 15, clip = true,width = 0.8\columnwidth]{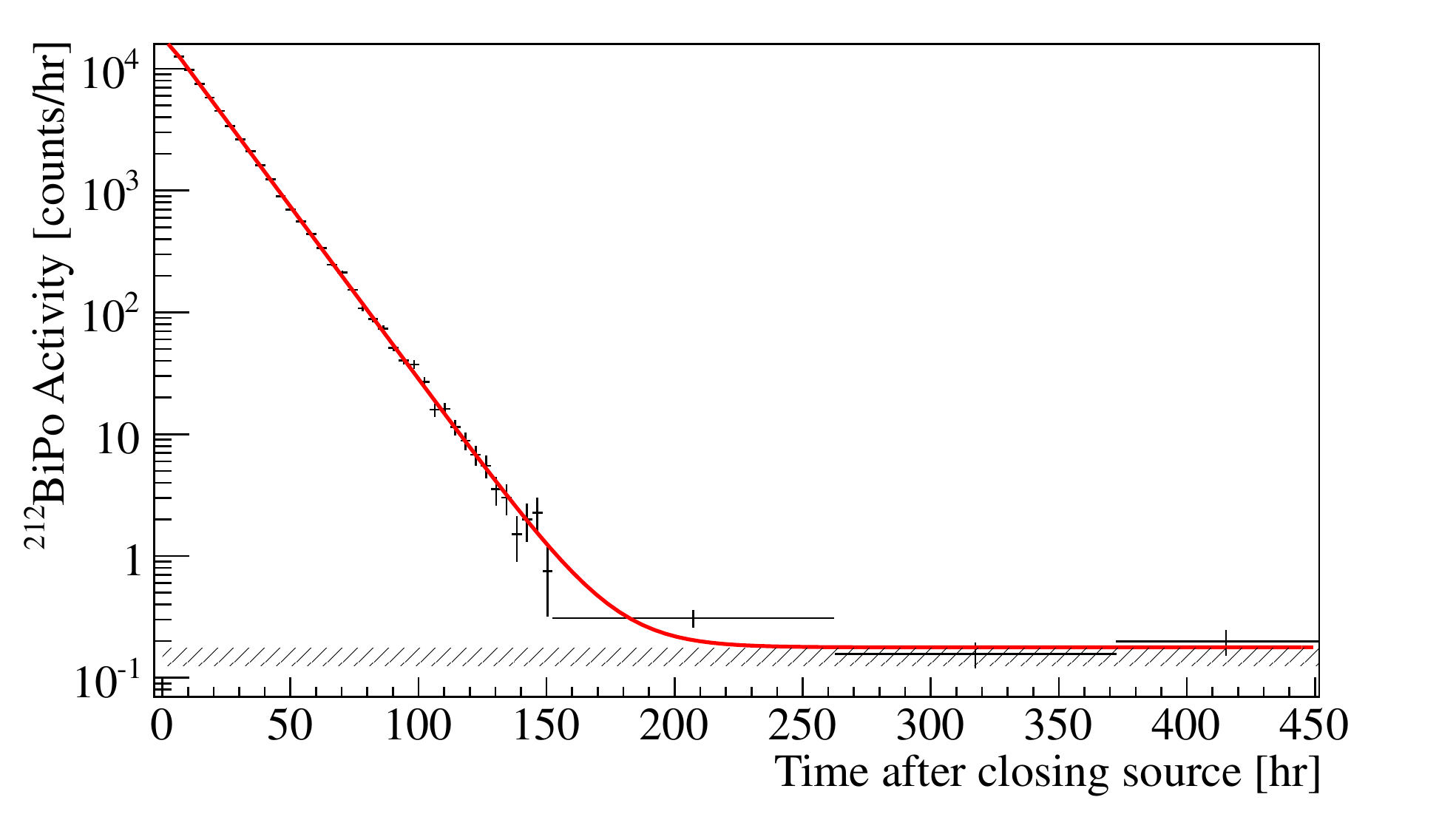}
\caption{Decay to background of $^{212}$BiPo in the Si PIN diode for the measurement of the sintered filter. An exponential and a constant are fitted to the data. The band is the previously measured background of the detector.}
\label{fig:bipo_background}
\end{figure}

The activity in the Si PIN diode was monitored as the released \Pb~decayed to background levels. A summary of the results are given in Table~\ref{tab:filter_limits}. About 14 half-lifes after closing the source the \Pb~has almost completely decayed to background levels, at which point any measurement would be sensitive to the release of \Ra. The background rate of $^{212}$BiPo events was found prior to these measurements to be $(42\pm7)\1{\mu Bq}$. Figure~\ref{fig:bipo_background} shows the decay for the measurement of the sintered filter, along with a fitted exponential and constant.

\begin{table}[htb]
\centering
\caption{Measurement results for the 0.5 micron sintered and ceramic filters}
\label{tab:filter_limits}
\renewcommand{\arraystretch}{1.2}
\begin{tabular}{|lccc|}
\hline\hline
Filter & Exposure & Fitted half-life & Fitted background \\ \hline
Sintered & $39\1{hr}$ & $(10.64\pm0.11)\1{hr}$ & $(49\pm8)\1{\mu Bq}$\\
Ceramic & $93\1{hr}$ & $(10.51\pm0.60)\1{hr}$ & $(46\pm8)\1{\mu Bq}$\\
\hline\hline
\end{tabular}
\end{table}

Both background values show some increase over the rate measured before this experiment, but in neither case is the increase statistically significant ($1\sigma$ for the sintered filter, $0.43\sigma$ for the ceramic filter). However, we can still (conservatively) attribute these small increases to some released \Ra, in which case we can limit the release of \Ra~to $<0.65\1{atoms/day/kBq}$ for the sintered filter and $<0.21\1{atoms/day/kBq}$ for the ceramic filter. Thus we see that both filters are highly effective at preventing the release of $^{224}$Ra from the source.

\begin{figure}[htb]
\centering
\includegraphics[trim = 5 0 40 20, clip = true,width = 0.8\columnwidth]{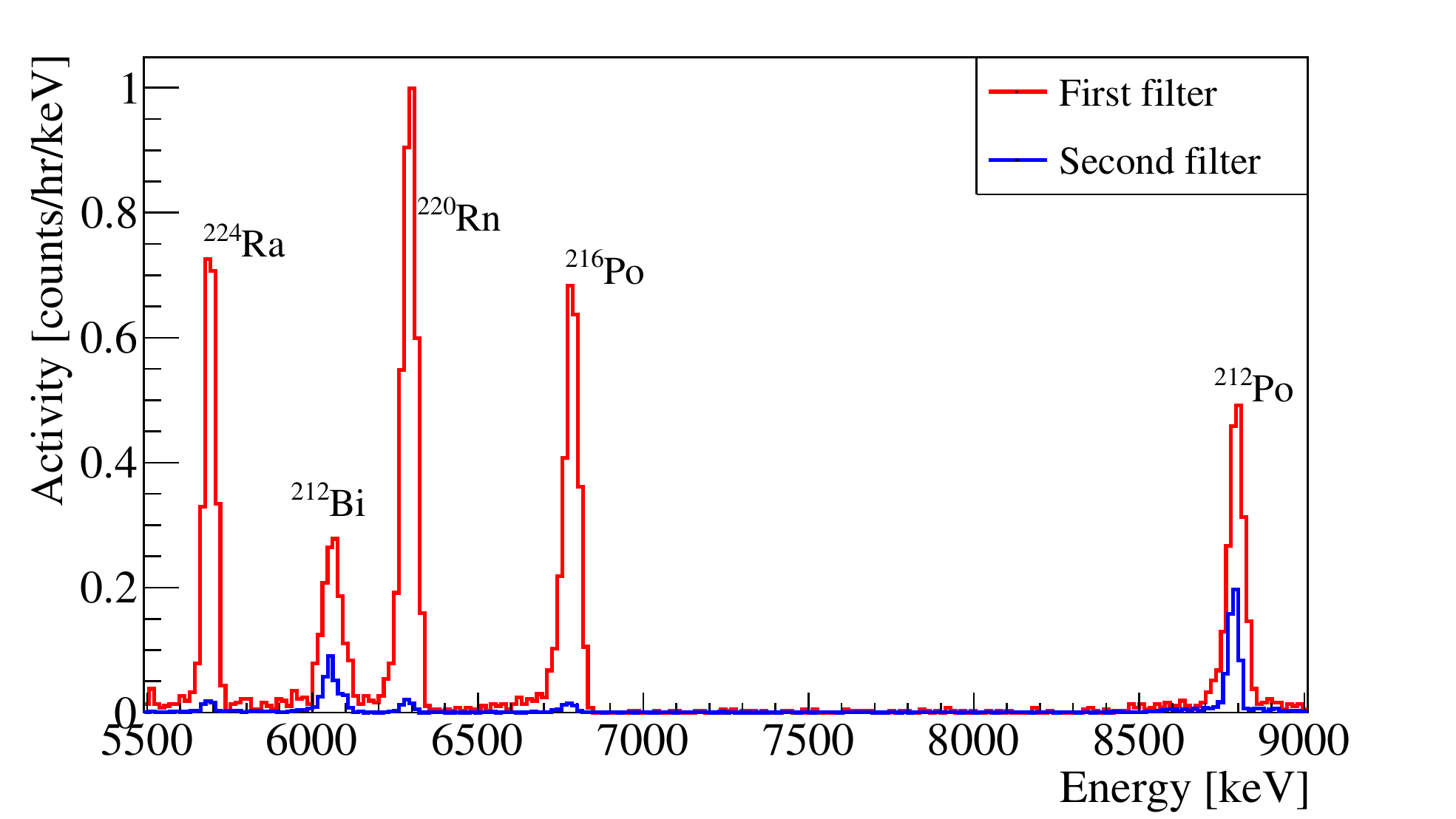}
\caption{Energy spectrum taken with the Si PIN diode from two sintered filters that were placed in series directly after the source in a xenon gas recirculation system.}
\label{fig:twofilters}
\end{figure}

A more direct measurement of filter efficiency was performed by placing two identical 90~micron sintered filters in series in the gas system immediately after the source vessel. After recirculating xenon gas at 6.5~slpm through the source and filters for 100~hours, the two filters were then placed on top of the Si PIN diode to measure any $\alpha$-activity coming off of them that could be attributed to \Ra. A total of 4~days of data were taken. The spectra of the two filters is shown in Figure~\ref{fig:twofilters}. While the first filter showed an activity of $(55.4\pm2.1)\1{mBq}$ of \Ra, the second filter that was placed immediately downstream of the first only showed an activity of $(1.63\pm0.18)\1{mBq}$ of \Ra. Both numbers are corrected for the activity at the time of source closing. Hence, their ratio directly gives the filter efficiency. This value is independent of systematic uncertainties such as the collection efficiency of the Si PIN diode, geometrical effects, etc. We thus find these 90~micron sintered filters to retain $(97.1\pm0.3)$\% of \Ra~flushing through them.

\section{Mixing in a Liquid Xenon Detector}

\begin{figure}[htb]
\centering
\includegraphics[trim = 0 0 0 0, clip = true,width = 0.8\columnwidth]{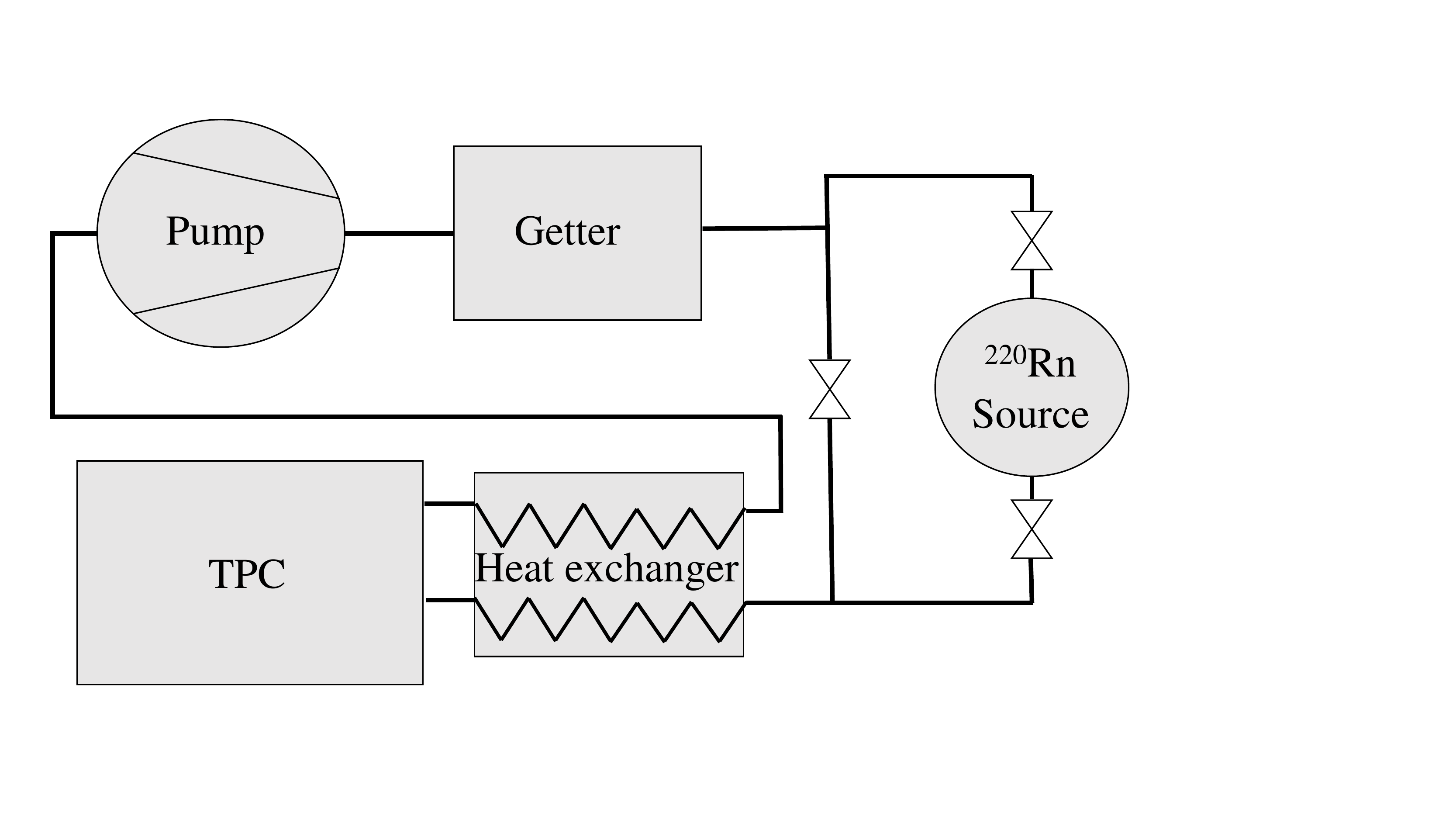}
\caption{Schematic of the recirculation loop for the liquid xenon detector measurement. The liquid is evaporated in the heat exchanger and is pumped through the purifier, where it can then be flushed past the \Rn~source or injected directly back to the detector.}
\label{fig:recirc}
\end{figure}

A liquid/gas xenon TPC was used to test the injection of \Rn~into a liquid xenon detector. The detector was designed with a long drift length of 17~cm and a diameter of 80~mm. Each end of the drift chamber is monitored by 7~Hamamatsu R8720 photomultiplier tubes (PMTs). The walls of the TPC are constructed of a single cylinder of PTFE, for use as a UV~light reflector, with field shaping electrodes embedded inside the PTFE walls. The drift and charge amplification fields are applied with transparent meshes at the ends of the TPC to allow for high optical collection. For the measurements presented here, only scintillation light was collected. However, this setup is sufficient to demonstrate that \Rn~can be flushed into the detector before decaying so that the subsequent \Pb~beta decays can be used as an internal calibrator. The detector was filled with about 3~kg of liquid xenon, which was continuously recirculated at 8.5~slpm through a high-temperature zirconium getter to remove electronegative impurities. The recirculation is done in the gas phase using 1/2" VCR piping with a heat exchanger in the xenon loop~\cite{Giboni:2011wx}, constituting a setup very similar to that of the XENON1T experiment~\cite{Aprile:2012jh}. The \Rn~source vessel was connected to the gas recirculation system such that the gas could be flushed past the source or bypass it. Figure \ref{fig:recirc} shows a schematic of the detector, purification loop, and \Rn~source. About 5~m of piping connected the \Rn~source to the TPC. For comparison, in XENON1T there are about 20~m of piping connecting the source to the TPC, and recirculation is planned for 100~slpm. Taken together, a comparable injection of activity may be expected.

\begin{figure}[htb]
\centering
\includegraphics[trim = 0 0 0 0, clip = true,width = 0.8\columnwidth]{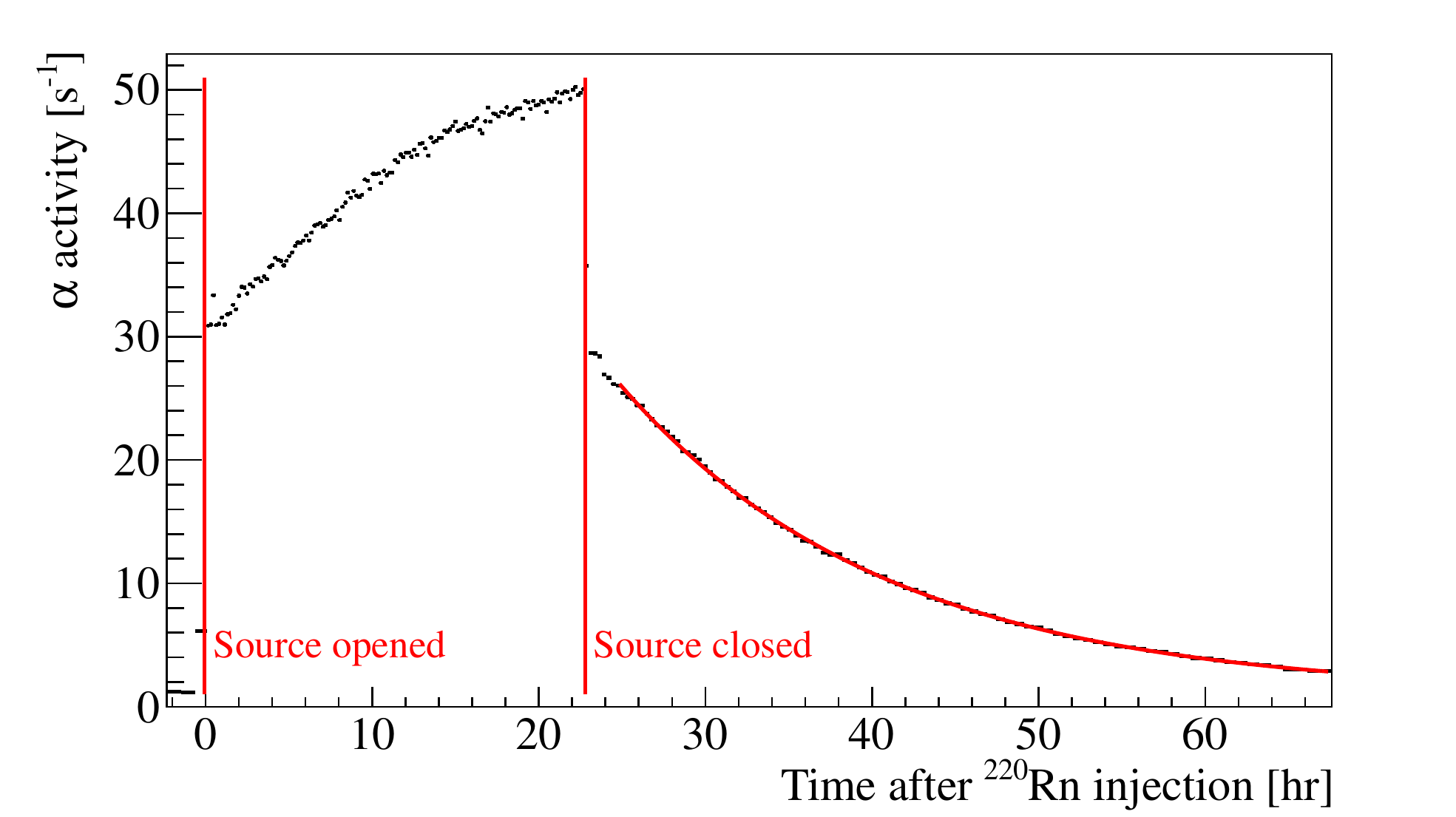}
\caption{Trigger rate of high-energy events in the liquid xenon detector before, during, and after source exposure. After source opening, the \Pb~activity grows in, clearly demonstrating the introduction of activity into the target volume. At source closing, the \Rn~and $^{216}$Po~alpha activity quickly decays away, leading to a drop in rate. Aterwards, the activity decays more slowly. A fit of an exponential plus constant background onto the decaying edge yields half-life ($11.18\pm0.04$)~hr. This is close to the \Pb~half-life, with the difference attributed to dead-time effects. Hence, this data demonstrates the presence of \Pb~decays in the TPC.}
\label{fig:Rates}
\end{figure}

To prove that the \Rn~could be mixed into the detector, the source was opened for 22.8 hours, injecting the doped gas. By monitoring the trigger rate of the detector, it was possible to observe the arrival of the dopants into the liquid xenon target. Data were acquired for 2.9~days surrounding this opening to determine the background before opening, and to monitor the decay of the injected \Pb~after the source was closed. The resulting trigger rate evolution of high-energy $\alpha$-decays is shown in Figure \ref{fig:Rates}. Events were selected by cutting at low energies and enforcing a coincidence requirement on the bottom PMTs. The background trigger rate of 1~Hz quickly rose to a 30~Hz after the source was opened, showing that the activity of the source was entering the liquid xenon target, followed by the growing in of \Pb~and its daughters over the next 22~hours. After the source was closed, the trigger rate quickly dropped to 30~Hz, after which it decayed toward the background rate.

The data acquisition system was not optimized for measuring high rates in a two-phase operation, resulting in a large rate-dependent dead time. This impedes our determination of the actual activity in the target volume, as well as our ability to accurately measure the decay rate following the closing of the source. However, this measurement is clearly sufficient to demonstrate three major features of this source: first, \Rn~emanates from the source and can be mixed into a liquid xenon detector. Second, a low activity source is capable of introducing enough activity to be measured even above a high background rate. Third, the decay after the source is closed demonstrates the presence of \Pb~in the liquid target, a requirement to use this source for calibration of the low-energy region in a dark matter search.

\section{Interpretation}

\begin{table}[htb]
\centering
	\caption{Summary of all measurements of emanation rates, given in units of atoms/min/kBq.}
	\label{tab:summary}
	\begin{tabular}{|lcc|}
		\hline \hline
		Measurement & \Th & \Ra \\ \hline
		$\gamma$ measurements of filters, Section~\ref{sec:tuv}  & $<34$ & $<0.66$ \\
		& $<0.4$ & $1.9\pm0.6$ \\
		Pipe contamination, Section~\ref{sec:flush} & $<47$ & $1.53\pm0.04$ \\
		Radon monitor, Section~\ref{sec:diode} & $<0.008$ & $3.9\pm1.3$ \\
		Si PIN diode, Section~\ref{sec:diode} & $270\pm12$ & $16926\pm6$ \\
		Sintered filter, Section~\ref{sec:filter} & N/A & $<4.5\times10^{-4}$ \\
		Ceramic filter, Section~\ref{sec:filter} & N/A & $<1.5\times10^{-4}$ \\
		\hline \hline
	\end{tabular}
\end{table}

Table~\ref{tab:summary} summarizes the various measurements. The two $\gamma$ measurements of filter deposition and the measurement of pipe contamination are methodologically very similar, yet the results are strikingly different. Additionally, the measurement using the Si PIN diode are orders of magnitude different. The apparent inconsistency between the $\gamma$ measurements and pipe contamination can be attributed to deposition of \Ra~in the 1/2" piping: in the first two measurements, 18~cm and 8~cm of piping were present respectively in between the source vessel and filter, while the copper pipe was less than 1~cm from the source (leaving just enough space to cold-weld the copper). In the second $\gamma$ measurement, the flow through the source vessel and the filter was laminar, whereas the flow was turbulent for the pipe contamination measurement. Due to its extremely high reactivity, radium will bond readily to pipe walls. This is facilitated by turbulent flow, while in laminar flow, the slow process of radial diffusion will impede the plate-out of \Ra. As the second $\gamma$ measurement and the pipe contamination measurement are consistent, we attribute the discrepancy of the first $\gamma$ measurement to the different conditions of the gas flow.

For the measurements of filter efficiency, the limits presented for the ceramic and 0.5~micron sintered filters are very satisfactory, while in the two-filter measurement, the 90~micron sintered filters are only 97\% efficient. However, at the time of this measurement, the source had an activity of around 60~kBq. Hence, the activity seen in the first filter ($55.4\pm2.1\1{mBq}$) shows a drastic reduction in the activity after only a few centimeters of piping. From this, we can conclude that even though the source gives off very large amounts of \Ra~and \Th~(as shown by the Si PIN diode measurement), only a vanishing percentage makes it out of the source vessel. With these sources being used in a fluid stream, one may thus expect the vast majority of \Ra~and \Th~to plate out in any connecting piping.

\section{Conclusions}

We have presented a versatile \Rn~source for the internal calibration of low-background detectors and demonstrated its suitability. Stray emanation was found to be $<0.008\1{atoms/min/kBq}$ for \Th~and $(1.53\pm0.04)\1{atoms/min/kBq}$ for \Ra, which can be reduced to $<10^{-3}\1{atoms/min/kBq}$ through the use of an additional filter. We have demonstrated that the \Rn~activity can be mixed under realistic conditions in a liquid noble gas detector. The source provides the means by which to calibrate the electronic recoil band in liquid noble element dark matter detectors at low energies, characterize the important radon backgrounds, map fluid dynamics in the liquid target and calibrate those detectors at energies above the Q-value of the $^{136}$Xe double-beta decay.

\section*{Acknowledgements}

This work was supported by National Science Foundation grants No.~PHY-1206061 and PHY-1209979, German Science Foundation DFG (WE1843/7-1), and by DFG Gro{\ss}ger\"at (INST 211/528-1 FUGG, funded together with the state NRW and University of M\"unster). We gratefully acknowledge support by Max-Planck-Gesellschaft as well as the Purdue Research Foundation.


\begin{thebibliography}{10}

\bibitem{Pandola:2014naa}
L.~Pandola, \emph{{Status of double beta decay experiments using isotopes other
  than $^{136}$Xe}},
  \href{http://dx.doi.org/10.1016/j.dark.2014.05.005}{\emph{Phys. Dark Univ.}
  {\bf 4} (2014) 17--22}, [\href{http://arxiv.org/abs/1403.3329}{{\tt
  1403.3329}}].

\bibitem{Undagoitia:2015gya}
T.~{Marrod\'an Undagoitia} and L.~Rauch, \emph{{Dark matter direct-detection
  experiments}},
  \href{http://dx.doi.org/10.1088/0954-3899/43/1/013001}{\emph{J. Phys.} {\bf
  G43} (2016) 013001}, [\href{http://arxiv.org/abs/1509.08767}{{\tt
  1509.08767}}].

\bibitem{Albert:2015ekt}
{\scshape EXO-200} collaboration, J.~B. Albert et~al., \emph{{Search for
  $2\nu\beta\beta$ decay of $^{136}$Xe to the 0$_1^+$ excited state of
  $^{136}$Ba with EXO-200}},  \href{http://arxiv.org/abs/1511.04770}{{\tt
  1511.04770}}.

\bibitem{Aprile:2013doa}
{\scshape XENON100} collaboration, E.~Aprile et~al., \emph{{Limits on
  spin-dependent WIMP-nucleon cross sections from 225 live days of XENON100
  data}},
  \href{http://dx.doi.org/10.1103/PhysRevLett.111.021301}{\emph{Phys.Rev.Lett.}
  {\bf 111} (2013) 021301}, [\href{http://arxiv.org/abs/1301.6620}{{\tt
  1301.6620}}].

\bibitem{Akerib:2015rjg}
{\scshape LUX} collaboration, D.~S. Akerib et~al., \emph{{Improved WIMP
  scattering limits from the LUX experiment}},
  \href{http://arxiv.org/abs/1512.03506}{{\tt 1512.03506}}.

\bibitem{Amaudruz:2014nsa}
P.~A. Amaudruz et~al., \emph{{DEAP-3600 Dark Matter Search}},  in
  \emph{International Conference on High Energy Physics 2014}, (Valencia,
  Spain), 7, 2014.
\newblock \href{http://arxiv.org/abs/1410.7673}{{\tt 1410.7673}}.

\bibitem{Calvo:2015uln}
{\scshape ArDM} collaboration, J.~Calvo et~al., \emph{{Status of ArDM-1t: First
  observations from operation with a full ton-scale liquid argon target}},
  \href{http://arxiv.org/abs/1505.02443}{{\tt 1505.02443}}.

\bibitem{Agnes:2015ftt}
{\scshape DarkSide} collaboration, P.~Agnes et~al., \emph{{Low Radioactivity
  Argon Dark Matter Search Results from the DarkSide-50 Experiment}},
  \href{http://arxiv.org/abs/1510.00702}{{\tt 1510.00702}}.

\bibitem{Aprile:2015uzo}
{\scshape XENON} collaboration, E.~Aprile et~al., \emph{{Physics reach of the
  XENON1T dark matter experiment}},
  \href{http://arxiv.org/abs/1512.07501}{{\tt 1512.07501}}.

\bibitem{Hannen:2011}
V.~Hannen, E.~Aprile, F.~Arneodo, L.~Baudis, M.~Beck, K.~Bokeloh et~al.,
  \emph{{Limits on the release of Rb isotopes from a zeolite based $^{83m}$Kr
  calibration source for the XENON project}},
  \href{http://dx.doi.org/10.1088/1748-0221/6/10/P10013}{\emph{Journal of
  Instrumentation} {\bf 6} (10, 2011) 10013},
  [\href{http://arxiv.org/abs/1109.4270}{{\tt 1109.4270}}].

\bibitem{Manalaysay:2009yq}
A.~Manalaysay, T.~{Marrod\'an Undagoitia}, A.~Askin, L.~Baudis, A.~Behrens,
  A.~D. Ferella et~al., \emph{{Spatially uniform calibration of a liquid xenon
  detector at low energies using $^{83m}$Kr}},
  \href{http://dx.doi.org/10.1063/1.3436636}{\emph{Review of Scientific
  Instruments} {\bf 81} (7, 2010) 073303},
  [\href{http://arxiv.org/abs/0908.0616}{{\tt 0908.0616}}].

\bibitem{Kastens:2010}
L.~W. Kastens, S.~Bedikian, S.~B. Cahn, A.~Manzur and D.~N. McKinsey, \emph{{A
  $^{83}$Kr$^m$ source for use in low-background liquid Xenon time projection
  chambers}},
  \href{http://dx.doi.org/10.1088/1748-0221/5/05/P05006}{\emph{Journal of
  Instrumentation} {\bf 5} (5, 2010) 5006},
  [\href{http://arxiv.org/abs/0912.2337}{{\tt 0912.2337}}].

\bibitem{Akerib:2013tjd}
{\scshape LUX} collaboration, D.~Akerib et~al., \emph{{First results from the
  LUX dark matter experiment at the Sanford Underground Research Facility}},
  \href{http://dx.doi.org/10.1103/PhysRevLett.112.091303}{\emph{Phys.Rev.Lett.}
  {\bf 112} (2014) 091303}, [\href{http://arxiv.org/abs/1310.8214}{{\tt
  1310.8214}}].

\bibitem{Akerib:2015wdi}
{\scshape LUX} collaboration, D.~S. Akerib et~al., \emph{{Tritium calibration
  of the LUX dark matter experiment}},
  \href{http://arxiv.org/abs/1512.03133}{{\tt 1512.03133}}.

\bibitem{WeberM:2013}
M.~Weber, \emph{{Gentle Neutron Signals and Noble Background in the XENON100
  Dark Matter Search Experiment}}.
\newblock PhD thesis, Ruprecht-Karls-Universit{\"a}t, Heidelberg, 7, 2013.

\bibitem{Albert:2015vma}
J.~Albert, D.~Auty, P.~Barbeau, D.~Beck, V.~Belov et~al., \emph{{Measurements
  of the ion fraction and mobility of alpha and beta decay products in liquid
  xenon using EXO-200}},  \href{http://arxiv.org/abs/1506.00317}{{\tt
  1506.00317}}.

\bibitem{Bellini:2012qg}
{\scshape Borexino} collaboration, G.~Bellini et~al., \emph{{Lifetime
  measurements of $^{214}$Po and $^{212}$Po with the CTF liquid scintillator
  detector at LNGS}},
  \href{http://dx.doi.org/10.1140/epja/i2013-13092-9}{\emph{Eur.Phys.J.} {\bf
  A49} (2013) 92}, [\href{http://arxiv.org/abs/1212.1332}{{\tt 1212.1332}}].

\bibitem{Auger:2012}
M.~Auger et~al., \emph{{The EXO-200 detector}},
  \href{http://dx.doi.org/10.1088/1748-0221/7/05/P05010}{\emph{JINST} {\bf 7}
  (2012) P05010}, [\href{http://arxiv.org/abs/1202.2192}{{\tt 1202.2192}}].

\bibitem{Baudis:2012}
{\scshape DARWIN} collaboration, L.~Baudis et~al., \emph{{DARWIN dark matter
  WIMP search with noble liquids}},
  \href{http://dx.doi.org/10.1088/1742-6596/375/1/012028}{\emph{Journal of
  Physics Conference Series} {\bf 375} (7, 2012) 012028},
  [\href{http://arxiv.org/abs/1201.2402}{{\tt 1201.2402}}].

\bibitem{Akerib:2015cja}
{\scshape LZ} collaboration, D.~S. Akerib et~al., \emph{{LUX-ZEPLIN (LZ)
  Conceptual Design Report}},  \href{http://arxiv.org/abs/1509.02910}{{\tt
  1509.02910}}.

\bibitem{Franco:2015pha}
D.~Franco et~al., \emph{{Solar neutrino detection in a large volume
  double-phase liquid argon experiment}},
  \href{http://arxiv.org/abs/1510.04196}{{\tt 1510.04196}}.

\bibitem{Budjas:1}
D.~Budj{\'a}{\v s} et~al., \emph{{Highly Sensitive Gamma-Spectrometers of GERDA
  for Material Screening: Part II}},  (Baksan, Russia), Proc 14th Intern.
  School, Particles and Cosmology, 4, 2007.
\newblock \href{http://arxiv.org/abs/0812.0768}{{\tt 0812.0768}}.

\bibitem{Budjas:2}
D.~Budj{\'a}{\v s}, M.~Heisel, W.~Maneschg and H.~Simgen, \emph{{Optimisation
  of the MC-model of a p-type Ge spectrometer for the purpose of effciency
  determination}}, {\emph{App. Rad. Isot.} {\bf 67} (2009) 706--710}.

\bibitem{Geant4}
S.~Agostinelli, J.~Allison, K.~Amako, J.~Apostolakis, H.~Araujo, P.~Arce
  et~al., \emph{{Geant4 - a simulation toolkit}},
  \href{http://dx.doi.org/http://dx.doi.org/10.1016/S0168-9002(03)01368-8}{\emph{Nuclear
  Instruments and Methods in Physics Research Section A: Accelerators,
  Spectrometers, Detectors and Associated Equipment} {\bf 506} (2003) 250 --
  303}.

\bibitem{zuzel_llrmt}
G.~Zuzel and H.~Simgen, \emph{High sensitivity radon emanation measurements},
  \href{http://dx.doi.org/http://dx.doi.org/10.1016/j.apradiso.2009.01.052}{\emph{Applied
  Radiation and Isotopes} {\bf 67} (2009) 889--893}.

\bibitem{Bray}
E.~P. Bray, R.~Lang and S.~MacMullin, ``{Characterizing New Calibration Sources
  in Liquid Xenon Dark Matter Searches}.'' (2014), available at \href{http://docs.lib.purdue.edu/surf/2014/presentations/18}{{\tt http://docs.lib.purdue.edu/surf/2014/presentations/18}}.

\bibitem{Firestone}
R.~B. Firestone and V.~S. Shirley, \emph{{Table of Isotopes}}.
\newblock John Wiley \& Sons, Inc., New York, 8~ed., 1998.

\bibitem{Giboni:2011wx}
K.~L. Giboni, E.~Aprile, B.~Choi, T.~Haruyama, R.~F. Lang, K.~E. Lim et~al.,
  \emph{{Xenon Recirculation-Purification with a Heat Exchanger}},
  \href{http://dx.doi.org/10.1088/1748-0221/6/03/P03002}{\emph{JINST} {\bf 6}
  (2011) P03002}, [\href{http://arxiv.org/abs/1103.0986}{{\tt 1103.0986}}].

\bibitem{Aprile:2012jh}
E.~Aprile et~al., \emph{{Performance of a cryogenic system prototype for the
  XENON1T Detector}},
  \href{http://dx.doi.org/10.1088/1748-0221/7/10/P10001}{\emph{JINST} {\bf 7}
  (2012) P10001}, [\href{http://arxiv.org/abs/1208.2001}{{\tt 1208.2001}}].

\end{thebibliography}

\providecommand{\href}[2]{#2}\begingroup\raggedright\endgroup

\end{document}